%% file: preprint.tex
\documentclass[aps,prb,a4paper,twocolumn,twoside,
unsortedaddress,eqsecnum,showpacs,showkeys,floatfix]{revtex4}
\usepackage{amsmath,amssymb,dcolumn,graphicx}
\input{macroses.tex}

\begin{document}
\title{Site-site memory equation approach in study of density/pressure dependence
of translational diffusion coefficient and rotational relaxation time of
polar molecular solutions:\newline
acetonitrile in water, methanol in water,  and methanol in acetonitrile}

\author{Alexander E. Kobryn}
\affiliation{Department of Theoretical Study, Institute for Molecular Science, Myodaiji, Okazaki, Aichi 444-8585, Japan}

\author{Tsuyoshi Yamaguchi}
\affiliation{Department of Molecular Design and Engineering, Graduate School of Engineering,\\
Nagoya University, Chikusa, Nagoya, Aichi 464-8603, Japan}

\author{Fumio Hirata}
\thanks{The author to whom correspondence should be sent.}
\affiliation{Department of Theoretical Study, Institute for Molecular Science,
and Department of Functional Molecular Science, The Graduate University for Advanced Studies,
Myodaiji, Okazaki, Aichi 444-8585, Japan}

\date{December 12, 2004}

\begin{abstract}
We present results of theoretical study and numerical calculation of the
dynamics of molecular liquids based on combination of the memory equation
formalism and the reference interaction site model -- RISM. Memory
equations for the site-site intermediate scattering functions are studied
in the mode-coupling approximation for the first order memory kernels,
while equilibrium properties such as site-site static structure factors
are deduced from RISM. The results include the
tem\-pe\-ra\-tu\-re-den\-si\-ty(pres\-su\-re) dependence of translational
diffusion coefficients $D$ and orientational relaxation times $\tau$ for
acetonitrile in water, methanol in water and methanol in acetonitrile,
all in the limit of infinite dilution. Calculations are performed over
the range of temperatures and densities employing the SPC/E model for
water and optimized site-site potentials for acetonitrile and methanol.
The theory is able to reproduce qualitatively all main features of
temperature and density dependences of $D$ and $\tau$ observed in real
and computer experiments. In particular, anomalous behavior, i.e. the
increase in mobility with density, is observed for $D$ and $\tau$ of
methanol in water, while acetonitrile in water and methanol in
acetonitrile do not show deviations from the ordinary behavior. The
variety exhibited by the different solute-solvent systems in the density
dependence of the mobility is interpreted in terms of the two competing
origins of friction, which interplay with each other as density
increases: the collisional and dielectric frictions which, respectively,
increase and decrease with increasing density.
\end{abstract}

\pacs{%
05.20.Jj -- Statistical Mechanics of Classical Fluids,
61.25.Em -- Molecular Liquids}

\keywords{%
Memory Equation,
Mode-Coupling,
RISM,
Translational Diffusion,
Rotational Relaxation
}

\maketitle

\input{section01.tex}

\input{section02.tex}

\input{section03.tex}

\input{section04.tex}

\input{section05.tex}

\input{section06.tex}

\newpage

\input{bibliography.tex}
\end{document}

%% file: macroses.tex
\newcommand*{\be}{\begin{equation}}
\newcommand*{\ee}{\end{equation}}
\newcommand*{\ba}{\begin{array}}
\newcommand*{\ea}{\end{array}}
\newcommand*{\bea}{\begin{eqnarray}}
\newcommand*{\eea}{\end{eqnarray}}
\newcommand*{\bean}{\begin{eqnarray*}}
\newcommand*{\eean}{\end{eqnarray*}}

\newcommand*{\lp}{\left(}
\newcommand*{\rp}{\right)}
\newcommand*{\ls}{\left[}
\newcommand*{\rs}{\right]}

\newcommand*{\la}{\langle}
\newcommand*{\La}{\left\la}
\newcommand*{\ra}{\rangle}
\newcommand*{\Ra}{\right\ra}

\def\tens#1{\mbox{\usefont{T1}{cmss}{bx}{n}#1}}

\renewcommand*{\d}{\textrm{d}}
\newcommand*{\e}{\mathrm{e}}
\newcommand*{\veps}{\varepsilon}

\newcommand*{\kB}{k_{\rm B}}
\newcommand*{\Tr}{\mathop{{\rm Tr}}\nolimits\,}
\newcommand*{\ds}{\displaystyle}

\newlength{\glength}
\newcommand*{\CPL}{Chem. Phys. Lett.\ }
\newcommand*{\JACS}{J. Am. Chem. Soc.\ }
\newcommand*{\JCP}{J. Chem. Phys.\ }
\newcommand*{\JML}{J. Mol. Liquids\ }
\newcommand*{\JPCM}{J. Phys.: Cond. Matt.\ }
\newcommand*{\JPC}{J. Phys. Chem.\ }
\newcommand*{\JPCA}{J. Phys. Chem. A\ }
\newcommand*{\JPCB}{J. Phys. Chem. B\ }
\newcommand*{\MP}{Mol. Phys.\ }

\newcommand*{\PRE}{Phys. Rev. E\ }
\newcommand*{\PRL}{Phys. Rev. Lett.\ }
\newcommand*{\PTP}{Progr. Theor. Phys.\ }

%% file: section01.tex
\section{Introduction}
\label{Section01}

The mobility of either polar molecules or small po\-ly\-ato\-mic ions in
dense dipolar liquids exhibits some ano\-ma\-li\-es that have attracted
considerable interest of scientists
\cite{Kell75,Easteal85,Wakai94,Wakai97,Wakai99,Harris97,Harris98,Harris99a,
Harris99b, Biswas95,Chakrabarti96,Nienhuys00,Sassi00,Netz02,Chowdhuri03}.
In this paper we pay attention to the study of the so-called ``anomalous
pressure behavior'' of polar binary molecular liquids by considering such
popular models as methanol in water, acetonitrile in water and methanol
in acetonitrile. In particular, we are interested in the
den\-si\-ty(pres\-su\-re) dependence of translational diffusion
coefficients and reorientation relaxation times of solutes. To begin
with, we would like to say first few words about anomalies as they are.

The very primitive (classical, not quantum) picture of a molecular liquid
could be the set of finite size structural particles -- molecules --
consisting of interacting sites, which are connected by chemical bonds.
Positions of molecules as well as their orientations are not fixed in
general and may vary randomly in time. The change may be achieved either
by translational or rotational motion. Intuitively, it is then clear to
understand that applying the external pressure for such system results in
making the intermolecular distances shorter, which as the consequence
will limit the molecular mobility. In this case one may expect a
monotonous decrease with pressure for such quantities as translational or
rotational diffusion coefficients. Limitation of the molecular mobility
can also be discussed in terms of some other dynamical characteristics
such as viscosity, dielectric relaxation, etc. This point of view has
been confirmed experimentally for numerous liquids and solutions
\cite{Vargaftik96}. At the same time,  some molecular liquids do not obey
strictly this abstraction and exhibit behavior which is different from
the intuitive picture. Ty\-pi\-cal representatives of such molecular
liquids are water and some aqueous solutions
\cite{Eisenberg69,Horne72,Franks72} whose mobility with pressure can be
enhanced. The enhancement of the molecular (or ionic) mobility under
pressure then has been identified as the anomalous one.

Historically, first attempts to explain dynamics of molecules in
solutions were based on the use of the so-called \textit{continuum} model
for solvents
\cite{Born20,Debye29,Onsager36,Boyd61,Zwanzig63a,Zwanzig63b}. In this
model the surrounding medium is regarded as a dielectric continuum
extending up to the surface of the particle. The particle itself is
treated as a rigid sphere of finite size with some charge for ions or
dipole moment for molecules. Dynamic behavior of the media are described
by the Debye equation in terms of the frequency dependent dielectric
susceptibility, and hydrodynamic effects are entirely neglected. The most
important contribution of the theory was that it succeeded in extending
the phenomenological approaches to consider the friction on a rotating
dipole. When a dipole is surrounded by a dielectric continuum, it
polarizes the medium. If the dipole rotates, the polarization should
relax so that it adjusts itself to the electric field produced by the new
orientation of the dipole. The relaxation process is associated with the
energy dissipation which is identified as the dielectric friction. Later
the theory was improved by obtaining velocity of fluid from linearized
Navier-Stokes equation for a viscous incompressible fluid with boundary
conditions of either perfect sticking or slipping at the surface of the
particle \cite{Zwanzig70a,Zwanzig70b,Zwanzig74}. Subsequently, there were
many another suggestions on improvements of the theory, e.g. the enhanced
``con\-ti\-nu\-um mechanical framework'' description has been given in
series of works \cite{Hubbard77,Hubbard78a,Hubbard78b}, where dissipation
function is constructed by adding the electrical dissipation to the
ordinary hydrodynamic one, etc. However, contrary to what was expected,
the continuum model turned out to be unable to explain properly numerous
contradictions of phenomenological laws with experimental data. One
example is the failure of the Stokes-Einstein relation in explaining the
dependence of  mobility of ions in polar liquids on their size. In order
to compromise the trouble it has been suggested to use the so-called
``effective'' radii of particles, or rewrite the relation so that it
appears in the fractional exponent form \cite{Zwanzig85}. Such kind of
revisions had some effects, but there were also other difficulties which
could not be resolved in principle. For instance, dielectric continuum
theories cannot explain observed difference in mobility of anions and
cations because the interaction between an ion and dielectric media does
not depend on the sign of the ion charge. That is enough to realize that
the insolvency of a conventional continuum model description can be
overcome only by a microscopic formulation of the problem.

The \textit{microscopic}, or statistical-mechanical, theory of
mo\-le\-cu\-lar liquids is now on its rise, and the considerable progress
has been achieved during last years both in the description of
equilibrium and non-equilibrium properties \cite{Hirata03}. It became
possible due to the aptly used reference interaction site model -- RISM
\cite{Hirata81,Hirata82,Hirata83} and its modifications, or in the case
of dynamics -- due to appropriate unification of RISM and generalized
Langevin equation (GLE) which is obtained in the Mori formalism
\cite{Mori65a,Mori65b}. The choice in favor of generalized Langevin
equation is not accidental. For example, an alternative way of describing
non-equilibrium processes can be based on the use of various types of
kinetic equations which are derived from the BBGKY hierarchy
\cite{Zubarev9697}. However, when it is applied to systems with high
density, there is an ambiguity in the construction of the collision
integral. Also the linearization procedure for the collision integral
cannot be justified by the same arguments which are used in the case of
systems with low density \cite{Ferziger72}. In Mori formalism such
questions do not appear. GLE describes the time evolution of several
dynamic variables in concern as a function of the representative point in
the phase space. All other variables which are not under explicit concern
are projected onto the dynamic variables of interest with the help of a
projection operator. The projection leads to an equation which looks
similar to the phenomenological Langevin equation containing the
frictional force proportional to the rate of change of the dynamic
variables, and the random force, which are related to each other by the
fluc\-tu\-a\-ti\-on-dis\-si\-pa\-ti\-on theorem. If one chooses as the
dynamic variables the density and the conjugated current of sites or
atoms of molecules in liquid, the theory gives a memory equation for the
dynamic structure factor of atoms, which describes the time evolution of
the site-site density pair correlation functions \cite{Hirata92}. Initial
values, i.e. site-site static structure factors and direct correlation
function which are necessary to solve this integro-differential equation,
are obtained from the RISM or one of its generalizations. A crucial
development of the theory is rather conceptual, not mathematical, in the
sense that it has provided a new concept to view dynamics of a molecule
in solution, which is quite different from the model traditionally
exploited in the field. The new model regards the molecular motion in
liquid as a correlated \textit{translational} motion of atoms: if two
atoms in a diatomic molecule are moving in the same direction, then the
molecule as a whole is making translational motion, while the molecule
should be rotating if its atoms  are moving in opposite directions. This
view differs from the traditionally developed rotational-translational
model which is based on the use of angular coordinates
\cite{Madden82,Calef89,Chandra89,Wei89,Wei90}. Another important issue of
RISM-based theories, which should be mentioned here, is that they
pioneered in explaining the observed difference in the behavior of anions
and cations in solution. Both equilibrium \cite{Hirata88} and
non-equilibrium \cite{Chong99} systems were studied, and now we know that
these differences should be attributed to dissimilarities of the
microscopic structure of solvent surrounding the ion.

The new theory of dynamic processes in molecular liquids has been under
development in last decade in works by Hirata and Chong
\cite{Hirata95,Chong03} and get its extension in studies by Hirata,
Yamaguchi and others
\cite{Yamaguchi01a,Yamaguchi01b,Yamaguchi02a,Yamaguchi02b,
Yamaguchi03a,Yamaguchi03b,Yamaguchi03c,Yamaguchi04a,Yamaguchi04b,Nishiyama03}.
Particularly, theoretical description of pressure dependence of the
molecular motion of liquid water and transport properties of ionic
liquids were subjects for investigation in recent papers by Yamaguchi
\textit{et al.} \cite{Yamaguchi03a,Yamaguchi03b}. In present manuscript
we continue this line and report our latest study which is the pressure
and temperature dependence of dynamics of polar molecules in solution.
The manuscript is organized as follows. In subsection \ref{SubSection02a}
we write down equations of motion, which are memory equations, for
site-site intermediate scattering functions, and give relations for
translational diffusion coefficient and reorientation correlation time in
terms of these functions. In subsection \ref{SubSection02b} we deal with
equilibrium properties of the system which we obtain using DRISM with
hypernetted chain (HNC) type of the closure. In order to solve a memory
equation one needs to set the model for memory kernels. This question is
discussed in some detail in subsection \ref{SubSection02c}. More
attention is paid to the so-called mode-coupling approximation, which is
used to study dynamical behavior of the system in the long-time scale. We
apply our formalism to investigate some realistic systems such as
acetonitrile in water, methanol in water and methanol in acetonitrile.
Set up for these models and details of numerical procedures in the
computational part of this work are briefly explained in section
\ref{Section03}. Obtained results and remedies are discussed in section
\ref{Section04}, and conclusions are given in section \ref{Section05}.

%% file: section02.tex
\section{Theory}

\subsection{Memory equations for site-site intermediate scattering functions}
\label{SubSection02a}

Time-correlation functions are convenient tools for de\-scri\-bing
properties of many-body systems and can di\-rect\-ly be associated with
experimental ob\-ser\-vab\-les. A ge\-ne\-ral theoretical scheme for
calculation of time-correlation functions is provided by the Mori
approach \cite{Mori65a,Mori65b}, where they are phrased in terms of
memory function equations. Hereafter we will follow the formalism by
Hirata and Chong \cite{Chong03}, which is the judicious unification of
the theory of dynamical processes in simple liquids based on the Mori
approach \cite{Mori65a,Mori65b}, and reference interaction site model for
molecular liquids \cite{Hirata81,Hirata82,Hirata83}. In this formalism
``slow'' variables of the system are partial number densities and
longitudinal current densities. Then, of practical interest are elements
of the matrix of the site-site intermediate scattering functions
$\tens{F}(k;t)$ and the matrix of their self-parts
$\tens{F}_{\mathrm{s}}(k;t)$ defined by
\begin{subequations}
\begin{eqnarray}
F^{\alpha\gamma}(k;t)&=&
\frac1N\la\rho^{\alpha,*}(\textbf{k};0)\rho^\gamma(\textbf{k};t)\ra,\\
F^{\alpha\gamma}_{\mathrm{s}}(k;t)&=&
\frac1N\la\rho^{\alpha,*}(\textbf{k};0)\rho^\gamma(\textbf{k};t)\ra_{\mathrm{s}},
\end{eqnarray}
\end{subequations}
where $N$ is the total number of particles, $\rho^\alpha(\textbf{k};t)$
is the site $\alpha$ number density in reciprocal space, $\textbf{k}$ is
the wave-vector, $k=|\textbf{k}|$, $t$ is time, $*$ means complex
conjugation, angular brackets denote appropriate statistical average
(e.g., canonical Gibbs ensemble average), and suffix ``s'' stands for
``self'' and means the correlations between two sites in a same molecule.
In the present paper, we consider the limit of infinite dilution, the
limit that the solute concentration vanishes. One has two types of memory
equations for the solvent subsystem and one for the solute subsystem
(indicated by the superscript ``u'') which are:
\begin{subequations}
\label{memeq}
\begin{align}
\ddot{\tens{F}}(k;t)&=-\la\boldsymbol{\omega}_k^2\ra\tens{F}(k;t)
-\int_0^t\d\tau\;\tens{K}(k;\tau)\dot{\tens{F}}(k;t-\tau),
\nonumber\\\label{mevc}\\
\ddot{\tens{F}}_{\mathrm{s}}(k;t)&=
-\la\boldsymbol{\omega}_k^2\ra_{\mathrm{s}}\tens{F}_{\mathrm{s}}(k;t)
-\int_0^t\d\tau\;\tens{K}_{\mathrm{s}}(k;\tau)\dot{\tens{F}}_{\mathrm{s}}
(k;t-\tau),\nonumber\\\label{mevs}\\
\ddot{\tens{F}}^{\mathrm{u}}_{\mathrm{s}}(k;t)&=
-\la\boldsymbol{\omega}_k^2\ra^{\mathrm{u}}_{\mathrm{s}}
\tens{F}^{\mathrm{u}}_{\mathrm{s}}(k;t)
-\int_0^t\d\tau\;\tens{K}^{\mathrm{u}}_{\mathrm{s}}(k;\tau)
\dot{\tens{F}}^{\mathrm{u}}_{\mathrm{s}}(k;t-\tau).
\nonumber\\\label{meus}
\end{align}
\end{subequations}
In these equations, dot over the quantity means its time derivative, and
the memory function matrices, denoted as $\tens{K}(k;t)$,
$\tens{K}_{\mathrm{s}}(k;t)$ and
$\tens{K}^{\mathrm{u}}_{\mathrm{s}}(k;t)$, describe the friction on the
motion of interaction sites and are subject for definition later.
Quantities $\la\boldsymbol{\omega}_k^2\ra$,
$\la\boldsymbol{\omega}_k^2\ra_{\mathrm{s}}$ and
$\la\boldsymbol{\omega}_k^2\ra_{\mathrm{s}}^{\mathrm{u}}$ are normalized
second order frequency matrices given by
\begin{subequations}
\bea
\la\boldsymbol{\omega}_k^2\ra&=&k^2\tens{J}_{\text{\sc l}}(k)\tens{S}^{-1}(k),\\
\la\boldsymbol{\omega}_k^2\ra_{\mathrm{s}}&=&k^2\tens{J}_{\text{\sc l}}(k)
\tens{S}_\mathrm{s}^{-1}(k),\\
\la\boldsymbol{\omega}_k^2\ra_{\mathrm{s}}^{\mathrm{u}}&=&
k^2\tens{J}_{\text{\sc l}}^{\mathrm{u}}(k)\tens{S}_\mathrm{s}^{\mathrm{u},-1}(k),
\eea
\end{subequations}
where $\tens{S}(k)\equiv\tens{F}(k;t=0)$,
$\tens{S}_{\mathrm{s}}(k)\equiv\tens{F}_{\mathrm{s}}(k;t=0)$ and
$\tens{S}_{\mathrm{s}}^{\mathrm{u}}(k)\equiv\tens{F}_{\mathrm{s}}^{\mathrm{u}}(k;t=0)$
are matrices of static site-site structure factors and their self parts,
respectively, while $\tens{J}_{\text{\sc l}}(k)$ is the matrix of static
longitudinal site-current correlation functions, i.e.
\bea
\lfloor\tens{J}_{\text{\sc l}}(k)\rfloor^{\alpha\gamma}=
\frac1N\sum_{i,j}\La v^{\alpha}_{i,z}v^{\gamma}_{j,z}
\e^{-i\textbf{k}\cdot(\textbf{r}_i^\alpha-\textbf{r}_j^\gamma)}\Ra,
\label{longitudinal-current}
\eea
where subscripts $i,j$ refer to molecules,
$\textbf{r}_i^\alpha\equiv\textbf{r}_i^\alpha(0)$ is the initial position
of site $\alpha$, and $v^{\alpha}_{i,z}\equiv{}v^{\alpha}_{i,z}(0)$ is
longitudinal component of the initial velocity of site $\alpha$. The
analytical expression for arbitrary shape of the molecule has been given
recently by Yamaguchi \textit{et al.} \cite{Yamaguchi04b} and reads
\bea
\lefteqn{\ds\lfloor\tens{J}_{\text{\sc l}}(k)\rfloor^{\alpha\gamma}=
\frac{\kB T}{M}j_0(k\ell^{\alpha\gamma})}\nonumber\\
&&{}+\frac{\kB T}3\ls j_0(k\ell^{\alpha\gamma})+j_2(k\ell^{\alpha\gamma})
\rs\nonumber\\
&&\qquad\qquad\times[\delta\textbf{r}^\alpha]^{\text{\sc t}}
\cdot\ls\Tr\tens{I}^{-1}\openone-\tens{I}^{-1}\rs
\cdot\delta\textbf{r}^\gamma\nonumber\\
&&{}-\frac{\kB T}{(\ell^{\alpha\gamma})^2}j_2(k\ell^{\alpha\gamma})
[\delta\textbf{r}^\alpha\times\delta\textbf{r}^\gamma]^{\text{\sc
t}}\cdot\tens{I}^{-1}
\cdot[\delta\textbf{r}^\alpha\times\delta\textbf{r}^\gamma],\nonumber\\
\eea
where $\kB$ is the Boltzmann constant, $T$ is thermodynamic temperature,
$M$ is the total mass of the molecule, $\ell^{\alpha\gamma}$ is the
distance between sites, $\delta\textbf{r}^\alpha$ is the vector pointed
from the center of mass to the site $\alpha$, $\tens{I}$ is tensor of
inertia moments of the molecule, $\openone$ is the diagonal unit matrix;
finally, $j_0$ and $j_2$ are spherical Bessel functions of the first
kind. Definition for $\tens{J}_{\text{\sc l}}^{\mathrm{u}}(k)$ is similar
to the one given by equation (\ref{longitudinal-current}) with the
difference that summation runs over the solute molecule only.

Expressions for static site-site structure factors and memory kernels are
discussed separately in sections \ref{Section03} and \ref{Section04},
respectively. Here we show how quantities of our interest in this paper,
which are solute's translational diffusion coefficient $D$ and rank-1
reorientation correlation time $\tau_{\mathrm{R}1}$, are related with
$\tens{F}_{\mathrm{s}}^{\mathrm{u}}(k;t)$. The diffusion coefficient can
be expressed in terms of the site-site velocity autocorrelation function,
while the relaxation time is integrated rank-1 reorientation
autocorrelation function. Both of them are described in terms of the
$\tens{F}_{\mathrm{s}}^{\mathrm{u}}(\textbf{k};t)$.

Following the derivation procedure presented in Ref.
\onlinecite{Chong98c}, the translational diffusion coefficient $D$ is
obtained~as
\bea
D&=&\frac13\int_0^\infty\d t\;Z^{\alpha\gamma}(t)\nonumber\\
&=&\frac13\int_0^\infty\d t\ls\frac1N
\sum_i\La\textbf{v}_i^\alpha(0)\cdot\textbf{v}_i^\gamma(t)\Ra_{\mathrm{s}}\rs\nonumber\\
&=&-\lim_{t\to\infty}\int_0^t\d\tau\lim_{k\to0}\frac1{k^2}
\lfloor\ddot{\tens{F}}_{\mathrm{s}}^{\mathrm{u}}(k;\tau)\rfloor^{\alpha\gamma},
\eea
where $Z^{\alpha\gamma}(t)$ is the site-site velocity autocorrelation
function with sites $\alpha$ and $\gamma$ belonging to the same molecule.

The rank-1 reorientation autocorrelation function
$C_{\boldsymbol{\mu}}(t)$ is defined by
\be
C_{\boldsymbol{\mu}}(t)=\frac{\sum_i\La\boldsymbol{\mu}_i(0)\boldsymbol{\mu}_i(t)\Ra}
{\sum_j\La|\boldsymbol{\mu}_j|^2\Ra},
\label{Cmu}
\ee
where $\boldsymbol{\mu}_i(t)$ is a vector fixed on the molecule $i$. In
our case it is the dipole moment and therefore can be described by the
linear combination of site coordinates as
\be
\boldsymbol{\mu}_i(t)=\sum_{\alpha}z_\alpha\textbf{r}^\alpha(t)
\label{mu}
\ee
with $z_\alpha$ being site partial charges. Since the molecule is
considered to be the neutral one, partial charges satisfy the condition
of electro-neutrality $\sum_{\alpha}z_\alpha=0$, and make definition of
$\boldsymbol{\mu}_i(t)$ invariant under the transformation of translation
of the laboratory reference frame. Putting equation (\ref{mu}) into
equation (\ref{Cmu}) and using properties of time-correlation functions
\cite{Berne70} one arrives at
\begin{subequations}
\bea
C_{\boldsymbol{\mu}}(t)&=&
\frac{\sum_i\sum_{\alpha\gamma}z_{\alpha}z_{\gamma}
\La\textbf{r}_i^\alpha(0)\textbf{r}_i^\gamma(t)\Ra}
{\sum_j\La|\boldsymbol{\mu}_j|^2\Ra},\\
\ddot{C}_{\boldsymbol{\mu}}(t)&=&
-\frac{N\sum_{\alpha\gamma}z_{\alpha}z_{\gamma}Z^{\alpha\gamma}(t)}
{\sum_j\La|\boldsymbol{\mu}_j|^2\Ra}.
\eea
\end{subequations}
Hence, the time development of both $Z^{\alpha\gamma}(t)$ and
$C_{\boldsymbol{\mu}}(t)$ is governed by the memory equation for the
self-part of the site-site intermediate scattering function
$\tens{F}_{\mathrm{s}}^{\mathrm{u}}(k;t)$.

In this work we restrict our consideration to reorientation relaxation
processes of the rank-1 only and hereafter will not use the subscript
$\mathrm{R}1$ in notations. For the (rank-1) reorientation relaxation
time we use definition \cite{Hansen86}
\be
\tau=\int_0^\infty\d t\;C_{\boldsymbol{\mu}}(t).
\ee

\vspace*{3ex}
\subsection{Equilibrium properties of molecular liquids via RISM/DRISM}
\label{SubSection02b}

Initial values of intermediate scattering functions, that we need to
solve memory equations (\ref{memeq}), can be obtained using the RISM
theory \cite{Hirata03,Hirata81,Hirata82,Hirata83}. It predicts static
structure of molecular fluids via the calculation of site-site pair
correlation functions. This method has been extensively used and proved
to be the powerful tool in the microscopic description of equilibrium
quantities of the system \cite{Hirata03}. In the present article, we
employ the DRISM version of the theory \cite{Perkyns92a,Perkyns92b}
instead of the standard RISM/HNC theory. The DRISM theory incorporates a
bridge function which is determined to reproduce the dielectric constant
of solvent in phenomenological fashion. The use of the DRISM theory is
known to improve the dielectric loss spectrum  of solvent quantitatively,
although it does not change the essential conclusion of the study with
respect to the pressure dependence of the dynamics of solute in solvent
\cite{Yamaguchi03c}.

\subsection{Mode-Coupling approximation for memory kernels}
\label{SubSection02c}

It makes sense to talk about translational diffusion coefficient and
reorientation relaxation time only when dynamics of the system is
considered in the long-time limit, i.e. when the time scale is large
enough for fast relaxation processes to be completed. Memory kernel of
the me\-mo\-ry equation in that case is usually constructed by the
so-called mode-coupling approximation \cite{Hansen86}. In
works by Chong \textit{et al.} \cite{Chong98b,Chong02} the conventional
mode-coupling theory has been extended to the case of molecular
li\-qu\-ids based on the interaction-site model. In particular,
mode-coupling expressions for memory function matrices, which we will
denote as $\tens{K}_{\text{\sc mc}}(k,t)$ and
$\tens{K}_{\mathrm{s},\text{\sc mc}}(k,t)$, were obtained as bilinear
combinations of site-site intermediate scattering functions or their self
parts as
\begin{widetext}
\begin{subequations}
\begin{eqnarray}
\lfloor\tens{J}_{\text{\sc l}}^{-1}(k)\cdot\tens{K}_{\text{\sc mc}}(k,t)\rfloor^{\alpha\gamma}
&=&\frac{\rho}{(2\pi)^3}\int\d\textbf{q}\;\Big\{ q_z^2\lfloor\tilde{\tens{c}}(q)
\cdot\tens{F}(q,t)\cdot\tilde{\tens{c}}(q)\rfloor^{\alpha\gamma}
\tens{F}^{\alpha\gamma}(|\textbf{k}-\textbf{q}|,t)\nonumber\\
&&{}-q_z(k-q_z)\lfloor\tilde{\tens{c}}(q)
\cdot\tens{F}(q,t)\rfloor^{\alpha\gamma}
\lfloor\tens{F}(|\textbf{k}-\textbf{q}|,t)
\cdot\tilde{\tens{c}}(|\textbf{k}-\textbf{q}|)\rfloor^{\alpha\gamma}\Big\},
\label{eq:MCT_corr}\\\relax
\lfloor\tens{J}_{\text{\sc l}}^{-1}(k)\cdot\tens{K}_{\mathrm{s},\text{\sc mc}}(k,t)\rfloor^{\alpha\gamma}
&=&\frac{\rho}{(2\pi)^3}\int\d\textbf{q}\;q_z^2\lfloor\tilde{\tens{c}}(q)
\cdot\tens{F}(q,t)\cdot\tilde{\tens{c}}(q)\rfloor^{\alpha\gamma}
\tens{F}_{\mathrm{s}}^{\alpha\gamma}(|\textbf{k}-\textbf{q}|,t),\label{eq:MCT_self}
\end{eqnarray}
\end{subequations}
here for simplicity the wave-vector $\textbf{k}$ has been taken to be
parallel to $z$-axis. It has been shown, however, that the proposed
expressions for memory functions underestimate friction in orientational
motions \cite{Yamaguchi02b}. According to the recipe by Yamaguchi and
Hirata \cite{Yamaguchi02b}, memory functions for the self-part should be
given by the linear combination of corresponding mode-coupling memory
functions as
\begin{eqnarray}
\left\lfloor\tens{K}_{\mathrm{s}}(k,t)\cdot\tens{J}_{\text{\sc l}}(k)\right\rfloor^{\alpha\gamma}
=\mathop{\sum_{m_{1,2,3}=\{x,y,z\}}}\limits_{\mu,\nu\in i}\relax
\left\langle u_{zm_1}^{(i)}Z_{m_1m_2}^{\alpha\mu}Z_{m_2m_3}^{\nu\gamma}u_{zm_3}^{(i)}
\e^{i\textbf{k}\cdot\left(\textbf{r}_i^\alpha-\textbf{r}_i^\mu
-\textbf{r}_i^\gamma+\textbf{r}_i^\nu\right)}\right\rangle
\lfloor\tens{J}_{\text{\sc l}}^{-1}(k)\cdot\tens{K}_{\mathrm{s},\text{\sc mc}}(k,t)\rfloor^{\mu\nu},
\label{mkvs}
\end{eqnarray}
where $u_{zm}^{(i)}$ stands for the unitary matrix that describes
rotation between molecular and space-fixed reference frames of the
molecule $i$, while $Z_{m_1m_2}^{\alpha\gamma}$ denotes
orientation-dependent site-site velocity correlation matrix in the
molecular coordinate system and is given by
\bea
Z_{m_1m_2}^{\alpha\gamma}&=&\frac{\kB T}{M}\delta_{m_1,m_2}
+\kB T[\textbf{e}_{m_1}\times\delta\textbf{r}^\alpha]^{\text{\sc t}}\cdot
\tens{I}^{-1}\cdot[\textbf{e}_{m_2}\times\delta\textbf{r}^\gamma].
\eea
Here $\textbf{e}_{m_1}$ is the unit vector in the $m_1$ direction. The
collective part of the memory function (neglecting the orientational
correlation between different molecules) is obtained as
\cite{Yamaguchi02b}
\begin{eqnarray}
\tens{K}(k,t)=\tens{K}_{\text{\sc mc}}(k,t)+\tens{K}_{\mathrm{s}}(k,t)
-\tens{K}_{\mathrm{s},\text{\sc mc}}(k,t).\label{mkvc}
\end{eqnarray}
In the case of solute corresponding quantities are given by relations
\begin{eqnarray}
\lfloor\tens{J}_{\text{\sc l}}^{\mathrm{u},-1}(k)\cdot\tens{K}_{\mathrm{s},\text{\sc mc}}^{\mathrm{u}}(k,t)\rfloor^{\alpha\gamma}
&=&\frac{\rho}{(2\pi)^3}\int\d\textbf{q}\;q_z^2\lfloor\tilde{\tens{c}}^{\mathrm{uv}}(q)
\cdot{\tens{F}}^{\mathrm{v}}(q,t)\cdot\tilde{\tens{c}}^{\mathrm{vu}}(q)\rfloor^{\alpha\gamma}
\tens{F}^{\mathrm{u},\alpha\gamma}_{\mathrm{s}}(|\textbf{k}-\textbf{q}|,t),\\
\lfloor\tens{K}^{\mathrm{u}}_{\mathrm{s}}(k,t)\cdot\tens{J}_{\text{\sc l}}^{\mathrm{u}}(k)\rfloor^{\alpha\gamma}
&=&\mathop{\sum_{m_{1,2,3}=\{x,y,z\}}}\limits_{\mu,\nu\in i}\relax
\left\langle u_{zm_1}^{(i)}Z_{m_1m_2}^{\alpha\mu}Z_{m_2m_3}^{\nu\gamma}
u_{zm_3}^{(i)}\e^{i\textbf{k}\cdot\lp\textbf{r}_i^\alpha-\textbf{r}_i^\mu
-\textbf{r}_i^\gamma+\textbf{r}_i^\nu\rp}\right\rangle
\lfloor\tens{J}_{\text{\sc l}}^{\mathrm{u},-1}(k)\cdot\tens{K}_{\mathrm{s},\text{\sc mc}}^{\mathrm{u}}(k,t)\rfloor^{\mu\nu}.\qquad
\end{eqnarray}
Here, superscript ``v'' is used to indicate the solvent subsystem, and
intermediate scattering functions $\tens{F}^{\mathrm{v}}(k,t)$ are those
obtained in the process of solving of memory equations (\ref{mevc}) and
(\ref{mevs}) with mode-coupling expressions for memory kernels deduced
from equations (\ref{mkvc}) and (\ref{mkvs}), respectively.
\end{widetext}

%% file: section03.tex
\section{Setup of models and numerical procedures}
\label{Section03}

We performed explicit calculations for several po\-pu\-lar systems,
namely, acetonitrile (CH$_3$CN) in water, methanol (CH$_3$OH) in water
and methanol in acetonitrile, all in the case of infinite dilution. As
for the structure and the intermolecular potential of water we employed a
model of the extended simple point charge (SPC/E) \cite{Berendsen87}. We
also put the Lennard-Jones (LJ) core on the hydrogen atoms in order to
avoid the undesired divergence of the solution of the RISM integral
equation. The LJ parameters of the hydrogen atom, the depth of the well
and the diameter, were chosen to be $0.046$ kcal/mol and $0.7$ \AA,
respectively.

In acetonitrile and methanol the methyl group was considered to be a
single interaction site (Me) located on the methyl carbon atom.
Consequently, all chemical compounds consist of three sites which
interact through the pair potential \cite{Edwards84,Jorgensen86}
\begin{equation}
\phi(r_{ij}^{\alpha\gamma})=
4\epsilon_{\alpha\gamma}\ls\lp\frac{\sigma_{\alpha\gamma}}
{r_{ij}^{\alpha\gamma}}\rp^{12}
-\lp\frac{\sigma_{\alpha\gamma}}{r_{ij}^{\alpha\gamma}}\rp^{6}\rs
+\frac{z_{\alpha}z_{\gamma}}{r_{ij}^{\alpha\gamma}},
\label{interaction-potential}
\end{equation}
i.e., LJ plus Coulomb. Here
$r_{ij}^{\alpha\gamma}=|\mathbf{r}_i^\alpha-\mathbf{r}_j^\gamma|$;
parameters $\epsilon_{\alpha\gamma}$ and $\sigma_{\alpha\gamma}$ are LJ
well-depths and LJ diameters defined as
$\epsilon_{\alpha\gamma}=\sqrt{\epsilon_\alpha\epsilon_\gamma}$ and
$\sigma_{\alpha\gamma}=(\sigma_\alpha+\sigma_\gamma)/2$, respectively,
with $\sigma_\alpha$ being the LJ diameter of a single site. Point
charges for acetonitrile were chosen to reproduce electrostatic potential
obtained in {\itshape ab initio} calculations \cite{Edwards84}. Numerical
values of masses of sites, parameters of the site-site interaction
potential (\ref{interaction-potential}) and cartesian coordinates of
sites are specified in Table~\ref{parameters1e}. Information about bond
length can be deduced from cartesian coordinates of sites $(x,y,z)$.
%
\input{table1.tex}

In calculations for acetonitrile or methanol in water the temperature of
the system was varied from $258.15$ to $373.15$ K, and the density of
water from $0.9$ to $1.2$ g/cm${}^3$; for the case of methanol in
acetonitrile the temperature of the system was varied from $293.15$ to
$323.15$ K, and the density of acetonitrile from $0.6726$ to $0.815$
g/cm${}^3$. Connection of the water parameters with thermodynamic
pressure is shown in Table~\ref{density-pressure} (except for the
metastable regions where we do not have reliable data).
\input{table2.tex}

Tem\-pe\-ra\-tu\-re/den\-si\-ty dependent dielectric constant
$\varepsilon$ for water used in numerical calculations has been evaluated
as a physical solution of an empirical nonlinear equation presented in
Ref. \onlinecite{LandoltBornstein80}:
\begin{equation}
\veps-\frac12\lp1+\frac1{\veps}\rp=\frac1v
\lp17+\frac{9.32\cdot10^4\lp1+\frac{153}{v\cdot{T}^{1/4}}\rp}{\lp1-3/v\rp^
2T}\rp,
\end{equation}
where $v$ is a molar volume in units of cm$^3$/mol, and $T$ is
thermodynamic temperature~in~K. This equation was also used in such
tem\-pe\-ra\-tu\-re/den\-si\-ty points where no experimental values
exist. Density and dielectric constant for acetonitrile at different
temperatures are indicated in Table~\ref{density-epsilon-acetonitrile}.
%
\input{table3.tex}

From the memory equation / mo\-de-co\-up\-ling theory and the DRISM/HNC
integral equation theory, the diffusion coefficients and the
reorientation relaxation times of solute molecules in solution can be
obtained based solely on the information about molecular shapes, inertia
parameters, intermolecular interaction potentials, temperature and
density. First, we calculate the site-site static structure factor by
solving the DRISM equation using the intermolecular interaction,
molecular shape, tem\-pe\-ra\-tu\-re and density. In order to improve the
convergence of the DRISM calculation, we use the method of the mo\-di\-fi\-ed
direct inversion in an iterative space (MDIIS) proposed by Kovalenko
\textit{et al.} \cite{Kovalenko99} From the static site-site structure
factor, we calculate the site-site intermediate scattering function using
the site-site memory equation with the mode-coupling approximation for
the memory kernels. The memory equation is time-integrated nu\-me\-ri\-cal\-ly.
Time-development of correlation functions in the $k\to0$ limit is treated
separately by the analytical li\-mi\-ting procedure of theoretical
expressions. In the numerical procedure, the reciprocal space is linearly
discretized as $k=(n+\frac12)\Delta k$, where $n$ is an integer from 0 to
$N_k-1$. Values of $\Delta k$ and $N_k$ are $0.061$ \AA${}^{-1}$ and
$2^9=512$, respectively. The choice for $N_k$ as the power of two is
coming as the requirement of the subroutine for the fast Fourier
transform, which is used in DRISM / MDIIS. The diffusion coefficient $D$
is calculated from the asymptotic slope of the time dependence of the
mean square displacement, and the orientational relaxation time $\tau$ is
deduced from the rotational autocorrelation function.

%% file: table1.tex
\begin{table}[!htb]
\fontencoding{T1}\fontfamily{ptm}\fontseries{m}\fontshape{n}\fontsize{8}{9.6}\selectfont
\caption{Masses of sites $m_\alpha$, parameters of the site-site interaction potential
(\ref{interaction-potential}) and cartesian coordinates of sites for water \cite{Berendsen87},
acetonitrile \cite{Edwards84} and methanol \cite{Jorgensen86}. All values of
$\epsilon_{\mathrm{LJ}}$ are multiplied by the factor $10^{14}$, a.u. stands for ``atomic units''.}
\label{parameters1e}
\begin{ruledtabular}
\begin{tabular}{c@{\hspace*{1em}}
	D{.}{.}{2.3}D{.}{.}{2.4}D{.}{.}{1.2}D{.}{.}{1.4}D{.}{.}{2.4}D{.}{.}{2.4}D{.}{.}{2.4}}
	&\multicolumn{1}{c}{$m_\alpha$}
	&\multicolumn{1}{c}{$z_\alpha$}
	&\multicolumn{1}{c}{$\sigma_\alpha$}
	&\multicolumn{1}{c}{$\epsilon_\alpha$}	&x	&y	&z\\
	&\multicolumn{1}{c}{(a.u.)}
	&\multicolumn{1}{c}{(a.u.)}
	&\multicolumn{1}{c}{(\AA)}
	&\multicolumn{1}{c}{(erg/molec)}
	&\multicolumn{1}{l}{(\AA)}
	&\multicolumn{1}{l}{(\AA)}
	&\multicolumn{1}{l}{(\AA)}\\\hline
O	&16.0	&-0.8476	&3.16	&1.084	&0	&0	&-0.0646\\
H	&1.008	&0.4238		&0.7	&0.3196	&0	&0.8165	&0.5127\\
H	&1.008	&0.4238		&0.7	&0.3196	&0	&-0.8165&0.5127\\
\\
Me	&15.024	&0.269		&3.6	&2.64	&0	&0	&1.46\\
C	&12.0	&0.129		&3.4	&0.6878	&0	&0	&0\\
N	&14.0	&-0.398		&3.3	&0.6878	&0	&0	&-1.17\\
\\
Me	&15.024	&0.265		&3.74	&1.4525	&-1.4246&0	&0\\
O	&16.0	&-0.7		&3.03	&1.1943	&0	&0	&0\\
H	&1.008	&0.435		&1.0	&0.3196	&0.3004	&0.8961	&0\\
\end{tabular}
\end{ruledtabular}
\end{table}

%% file: table2.tex
\begin{table}[!htb]
\fontencoding{T1}\fontfamily{ptm}\fontseries{m}\fontshape{n}\fontsize{8}{9.6}\selectfont
\caption{Density-pressure correspondence for water \cite{Wagner02}
given for temperatures $T_1=273.15$ K, $T_2=298.15$ K and $T_3=373.15$ K.}
\label{density-pressure}
\begin{ruledtabular}
\begin{tabular}{D{.}{.}{1.3}D{.}{.}{1.5}D{.}{.}{3.4}D{.}{.}{3.4}D{.}{.}{3.4}}
	\multicolumn{1}{c}{Density}
	&\multicolumn{1}{c}{Number density}
	&\multicolumn{1}{c}{Pressure at $T_1$}
	&\multicolumn{1}{c}{Pressure at $T_2$}
	&\multicolumn{1}{c}{Pressure at $T_3$}\\
	\multicolumn{1}{c}{(g/cm$^3$)}
	&\multicolumn{1}{c}{(\AA$^{-3}$)}
	&\multicolumn{1}{c}{(MPa)}
	&\multicolumn{1}{c}{(MPa)}
	&\multicolumn{1}{c}{(MPa)}\\\hline
0.900	&0.03008&-&-&-\\
1.000	&0.03334&0.4085		&6.6914		&100.6450\\
1.025	&0.03426&52.6119	&66.3549	&173.0629\\
1.050	&0.03510&111.7871	&133.7116	&255.2558\\
1.075	&0.03593&179.4861	&209.9492	&347.8493\\
1.100	&0.03676&257.2009	&296.1955	&451.5134\\
1.125	&0.03760&346.1457	&393.4575	&566.9920\\
1.200	&0.04011&689.0266	&760.7554	&993.3837\\
\end{tabular}
\end{ruledtabular}
\end{table}

%% file: table3.tex
\begin{table}[!htb]
\fontsize{8}{9.6}\selectfont
\caption{Density and dielectric constant for acetonitrile as functions of temperature:
experimental data used in our computation. Temperature is in K, and density is in g/cm$^3$.}
\label{density-epsilon-acetonitrile}
\begin{ruledtabular}
\begin{tabular}{lccccccc}
$T$&\;\;293.13&\;\;295.05&\;\;298.15&\;\;303.15&\;\;308.15&\;\;313.15&\;\;323.15\\\hline
$\rho$&\;.782$^{\mathrm{a}}$&---&\;.7762$^{\mathrm{b}}$&\;.7712$^{\mathrm{a}}$&\;.7652$^{\mathrm{b}}$&\;.7603$^{\mathrm{a}}$&\;.7492$^{\mathrm{a}}$\\
$\veps$&\;38.8$^{\mathrm{c}}$&\;37.5$^{\mathrm{d}}$&\;36.69$^{\mathrm{e}}$&\;35.93$^{\mathrm{e}}$&&&\;33.5$^{\mathrm{f}}$\\
\end{tabular}
\end{ruledtabular}
$^{\mathrm{a}}$Ref. \onlinecite{Ku98}, 
$^{\mathrm{b}}$Ref. \onlinecite{Nikam98},
$^{\mathrm{c}}$Ref. \onlinecite{Delta},
$^{\mathrm{d}}$Ref. \onlinecite{ASI},
$^{\mathrm{e}}$Ref. \onlinecite{Cavell65},
$^{\mathrm{f}}$Extrapolated.
\end{table}

%% file: section04.tex
\section{Results and discussions}
\label{Section04}

\subsection{Equilibrium properties: static structure}

\input{figure1.tex}

\input{figure2.tex}

\input{figure3.tex}

\input{figure4.tex}

Since the discovery of anomalous behavior of the mo\-le\-cu\-lar mobility
with pressure it has been attributed to the hydrogen bonding properties
of the system. The idea is that the hydrogen bonds or hydrogen bonding
network are distorted upon compression, so that the dissolved molecule is
actually having fewer hydrogen bonds at higher pressure, thus making it
easier to move and rotate. The extensive test of this idea has been the
objective of many  researches. In particular, the time evolution of the
formation and rupture of hydrogen bonds turns out to be a popular
quantity to be examined in studies based on the computer simulation (see,
e.g., Ref. \onlinecite{Luzar00} and references therein). In order to
circumvent the problem,  a number of alternative  definitions  of
hydrogen bonds was suggested on top of the traditional one based either
on geometric \cite{Mezei81} and/or energetic \cite{Rahman71} criteria,
though none of them is conclusive. The definition of hydrogen bond in the
theoretical study based on the integral equation is somewhat ambiguous as
well. However, it is a quite general observation for such liquids like
water and alcohol that a pair of atoms having small size and/or large
charges with opposite sign makes a well-defined first peak in the pair
correlation function: for example, hydrogen and oxygen in water and
alcohol. Especially, in order for making a hydrogen bond, one of the pair
of atoms should be as small as a hydrogen atom. Since those li\-qu\-ids
are known to make a hydrogen bond between the pair of atoms, the
well-defined peak has been assigned to a ``hydrogen bond''. The hydrogen
bond between a pair of atoms induces a ``liquid structure'' which is
specific to molecules consisting the fluid, which in turn produces
peculiar peaks in the pair correlation function at par\-ti\-cu\-lar
distances. For example, in the case of water, the hydrogen bond peak
appears at $\sim1.8$ \AA{} in the O-H correlation function, which induces
a new peak in the O-O correlation function around $\sim4.5$ \AA. The O-O
distance is peculiar to the ice-like tetrahedral coordination. Therefore,
the peak at $\sim4.5$ \AA{} is regarded as a ``finger print'' of  the
hydrogen bond network existing in liquid water \cite{Eisenberg69}.
Alcohol also makes a hydrogen bond between a pair of the molecules, but
does not make a three dimensional network, which is distinct from water.
Instead, it makes zigzag chains of hydrogen bonds. It is because one of
the four hydrogen-bonding sites is replaced by an alkyl group which is
too large to form the bond. In the case of neat acetonitrile, there is no
indication for making hydrogen bonds between a pair of the molecules. It
is understood from its molecular constituents, i.e. it  does not have a
small atom as a possible hydrogen-bonding site. Therefore, the structure
of the liquid or the pair correlation function is determined essentially
by the repulsive interactions and the molecular geometry.

The pressure or density dependence of the liquid structure is determined
essentially by the interplay between the repulsive interaction and the
hydrogen bonding, which interfere each other through the intramolecular
constraints or molecular geometry. Increasing pressure will make
repulsive interactions more dominant, which will weaken the hydrogen bond
by distorting the fa\-vo\-ra\-ble arrangement between the two molecules.
The pair correlation functions on pressure or density should reflect the
interplay mentioned above. Increasing pressure or density will result in
reduction in all the indications of hydrogen bonding and of the
consequential liquid structure, and will make the steric or packing
effect more do\-mi\-nant.

Fig. \ref{gww} shows the site-site radial distribution functions (RDF's)
for neat water. The sharp peak at $\sim1.8$ \AA{} in the O-H RDF, Fig.
\ref{gww}b), is that of the hydrogen bond without any question. A weak
but conspicuous second peak at $\sim4.5$ \AA{} in the O-O RDF, Fig.
\ref{gww}a), is indicative of the tetrahedral configuration
characteristic in the hydrogen-bond network in water. All the changes in
the RDF caused by increasing density  coincide with the intuitive picture
emerged from the disruption or distortion of the hydrogen bond and its
network: the reduction of the hydrogen bond peak in the O-H RDF, the
reduction of the second peak in the O-O RDF. The increasing height in the
first and third peaks in the O-O RDF implies the enhanced tendency of the
steric and packing effect due to the increased density.

Fig. \ref{gwm} shows the site-site RDF for methanol in water. There are
two possible hydrogen bond sites in methanol, O and H, which make
hydrogen bonds, respectively with H and O of water molecules. The
O(water)-H(methanol) RDF, Fig. \ref{gwm}b), has the hydrogen bond peak at
the position where that in neat water appears. The peak height and
resolution are less compared to the case in neat water, implying the
hydrogen bond is somewhat weaker. The density dependence of the peak
shows si\-mi\-lar behavior with that  in neat water, indicating that the
hydrogen bond is weakened due to the steric distortion caused by
pressure. The O(methanol)-H(water) RDF, Fig. \ref{gwm}c), has the
hydrogen bond peak at $\sim1.5$ \AA, which is much shorter than that in
the O(water)-H(methanol) RDF. The peak is even more sharply defined than
that in neat water, implying that the hydrogen bond is stronger. The
density dependence of the peak shows similar tendency with that in neat
water, but decreases more sharply with increasing density. The
O(water)-O(methanol) RDF, Fig. \ref{gww}a), maintains the essential
structure which is seen in that of neat water. The important point is
that the second peak at $\sim4.5$ \AA{} has not disappeared, but is even
enhanced, which indicates the tetrahedral coordination peculiar to the
hydrogen bond network in water keeps its integrity. Indeed, in our
earlier work concerning the structure of the tert-butyl alcohol-water
mixture, we have clarified that a butanol molecule is incorporated in the
hydrogen-bond network of water in its infinite dilution by forming
hydrogen bonds with surrounding solvent molecules \cite{Yoshida02}.

The site-site RDF of the acetonitrile in water is shown in Fig. \ref{gwa}
along with its density dependence. Ac\-cor\-ding to the criteria for
hydrogen bonding ability of atoms stated above, the acetonitrile molecule
has one possible hydrogen bond site, nitrogen N, which can act only as a
proton acceptor. As is anticipated from the argument, the
N(acetonitrile)-H(water) RDF has a conspicuous hydrogen bond peak at
about $\sim1.8$ \AA, which again decreases with increasing density, Fig.
\ref{gwa}b). On the other hand, the N(acetonitrile)-O(water) RDF shows
less conspicuous second peak at 4.5 \AA, which implies that the hydrogen
bond network of water is somewhat disturbed around the solute compared to
the neat solvent, Fig. \ref{gwa}a).

In Fig. \ref{gam}, shown  is the site-site RDF for methanol in
acetonitrile at infinite dilution. In this case, a solvent (acetonitrile)
molecule can make a hydrogen bond with the solute molecule only at the
N-site by accepting the proton from methanol. The first peak in Fig.
\ref{gam}b) is indicative of the hydrogen bond. The solvent  molecules do
not make any hydrogen bond among themselves. The N-site never forms a
hydrogen bond with the oxygen site of methanol. It is not only because
the both sites have ne\-ga\-ti\-ve charges, but also because the neither
of them is as small as hydrogen. Therefore, the
N(acetonitrile)-O(methanol) RDF is largely determined by the repulsive
core and geometric constraints. The system is a good example of the case
in which constituent molecules can make only one hydrogen bond, only
between solute and solvent.

\subsection{Dynamical properties: diffusion coefficient and reorientation relaxation time}

Fig. \ref{diffusion} shows the density(pressure) dependence of the
normalized translational diffusion coefficient $D/D_0$ for investigated
solutes at various temperatures. Here $D_0$ is the translational
diffusion coefficient of the solute at the ambient density for the
solvent, which is $\rho=0.997047$ g/cm$^3$ for water in cases a), b), and
$\rho=0.782$ g/cm$^3$ for acetonitrile in the case c), respectively. Fig.
\ref{diffusion}a) shows that for sufficiently low temperature the
diffusion coefficient of methanol in water first increases with
density(pressure), and then begins to decrease. A relatively flat maximum
is observed for the density slightly smaller than that at the ambient
condition. The increase in the diffusion coefficient with increasing
density (or pressure) can not be explained by the phenomenological theory
such as the Stokes-Einstein law. In that reason, such behavior is often
stated ``anomalous''. This is the first to realize such anomalous
pressure dependence of the diffusion constant of a solute in water by
means of the statistical mechanics theory. As can be seen in the figure,
the anomalous density dependence is suppressed as temperature increases
and turns into normal behavior, i.e., the diffusion constant decreases
monotonously with density. At the ambient condition, the diffusion
constant does not show any anomaly and is consistent with the one
obtained from the molecular dynamics simulation \cite{Chowdhuri03}, which
is plotted in the figure with the filled triangles. Using the
multi-parametric empirical equation of state for water \cite{Wagner02} we
converted original data \cite{Chowdhuri03} from ``diffusion versus
pressure'' into ``diffusion versus density''.

Similar features can be observed in the behavior of the diffusion
coefficient of acetonitrile in water, Fig. \ref{diffusion}b). In this
case, the increase of the diffusion coefficient at the lowest temperature
is discernible, but very small as can be barely observed with the guide
of the cubic spline curve. Behavior of the diffusion coefficient at
higher temperatures, as in the case of methanol in water, does not show
any anomaly. The behavior is in harmony with the results of the molecular
dynamics simulation at the ambient condition \cite{Chowdhuri03}, which is
plotted in the figure with the filled squares. The experimental
measurements of diffusion coefficients for acetonitrile in water and
acetonitrile-D$_3$ (CD$_3$CN) in water, both at $T=303$ K, reported by
Nakahara \textit{et al.} \cite{Wakai94} testify the same tendency in
their density(pressure) dependence as those obtained from the theory.

The diffusion coefficient of methanol in acetonitrile are plotted in Fig.
\ref{diffusion}c) as the function of density. The results show monotonous
decrease of the diffusion coefficient in the entire range of densities
and at all investigated temperatures.
\input{figure5.tex}

\input{figure6.tex}

\input{figure7.tex}

Fig. \ref{relaxationtime} shows the density(pressure) dependence of the
normalized reorientation relaxation time $\tau/\tau_0$ for investigated
systems at the same temperatures as for normalized translational
diffusion coefficient $D/D_0$. $\tau_0$ is the reorientation relaxation
time of the solute at the ambient density for the solvent. Fig.
\ref{relaxationtime}a) is a typical example of anomalous
density(pressure) dependence of the reorientation relaxation time. It is
clear to see that it first decreases with density(pressure) and then
starts to increase. The anomalous behavior becomes weaker as temperature
increases, and disappears at some point. Apparently, the anomaly persists
through higher temperature than what is observed in the diffusion
coefficient. The theoretical result is consistent again with that from
the molecular dynamics simulation at the ambient condition
\cite{Chowdhuri03} in the sense that the both plots show a mi\-ni\-mum,
although the positions of the minimum are slightly different from each
other.

In the case of acetonitrile in water, Fig. \ref{relaxationtime}b), the
normal behavior of the reorientation relaxation time is observed almost
entirely except for the lowest temperature, where a shallow minimum is
discernible. The relaxation time at the ambient temperature exhibits
monotonous increase. The behavior is in qualitative accord with the
results from molecular dynamics simulation \cite{Chowdhuri03} as
indicated with the filled squares in the figure.

Fig. \ref{relaxationtime}c) exhibits monotonous increase of the
reorientation relaxation time of methanol in acetonitrile in the entire
range of densities and at all investigated temperatures. It can be
regarded as a typical example of normal density(pressure) dependence of
$\tau$.

Fig. \ref{relaxationfunction} has the purpose to demonstrate the
anomalous density(pressure) behavior of the reorientation relaxation time
in terms of the reorientation time-correlation function
$C_{\boldsymbol{\mu}}(t)$. It has been found that at $T=298.15$ K the
curves for acetonitrile in water and methanol in acetonitrile move to the
right as density increases, namely the relaxation time becomes longer
when pressure is increased. On the other hand, for methanol in water it
shifts first to the left, and then for higher densities(pressures) it
starts to shift to the right. At the highest temperature
$C_{\boldsymbol{\mu}}(t)$ for methanol in water moves only to the right,
but the displacement is very small. Considering that the time-correlation
function is related to the reorientation time by an integral, the
behaviors in the time-correlation functions are consistent with those of
the reorientation time as has been described above.

In our previous studies concerning the neat water, we have clarified the
cause of anomalous dependence of the molecular dynamics on pressure, or
the appearance of extrema in the pressure dependence
\cite{Yamaguchi03a,Yamaguchi04a}. The anomalous behavior is caused by an
interplay of two competing tendencies: the increasing mobility at the
lower pressure and the suppression of the motion in the higher pressure.
The suppression of the motion with in\-cre\-a\-sing pressure is a normal
behavior caused by increased molecular collisions, and the increase in
the mobility is caused by the decrease in the dielectric
relaxation time. The appearance
of the extrema is attributed to the increasing dielectric motion at the lower
pressure. The origin of the decrease in the dielectric relaxation time
at the lower pressure is in the reduction of the electrostatic
friction on the dielectric mode.
Although the detailed analysis is presented in our previous paper,
let us summarize it below.
The dielectric mode describes the collective alignment of the polar
molecules.  When the inhomogeneity of the number density is present as is
described in Fig. \ref{efriction}a),
\input{figure8.tex}

the alignment of the collective dipole moment induces the inhomogeneity
of the charge density, which leads to the electrostatic friction on
the dielectric mode. Increasing density or
pressure will suppress the number density fluctuation, which in turn
causes the reduction of the charge density fluctuation as is illustrated
in Fig. \ref{efriction}b). The reduced charge density fluctuation in turn
causes decrease in the electrostatic friction on the dielectric mode,
or increase in the collective rotational
motion. This is the rough picture with respect to the anomalous pressure
dependence of the rotational dynamics in water, which is consistent with
the analysis based on the mode coupling theory \cite{Yamaguchi03a}. 

Now we turn to the single-particle motion in water.
In the case of neat water, the translational and the rotational diffusion
of a molecule respond to pressure differently, as is observed experimentally.
The translational motion of a molecule couples with the high wave vector
collective translational
mode, which have inherent tendency to become slower as pressure increases.
On the other hand, the coupling between the rotational motion of a molecule
and high wave vector collective translational modes is weak, and the friction
on the rotation of a molecule mainly comes from the coupling with the
low wave vector charge density mode, which is usually referred to as the
dielectric friction.
The rotational motion of a molecule will induce a relaxation process of 
surrounding molecules in order
to make themselves aligned to the electric field produced by the new
orientation of the molecule in concern. The energy dissipation associated
with the relaxation process is an origin of the friction on the
rotational motion, or the dielectric friction. Slower the dielectric relaxation,
greater the dielectric friction.  The
anomalous pressure dependence observed in the translational dynamics or
the diffusion coefficient turns out to be the secondary effect induced by
the anomaly in the rotational motion.  The anomaly in the translational
diffusion is therefore weaker than that in the rotational relaxation.
\enlargethispage{4ex}

The solute-solvent systems considered in the present paper can be
understood in a similar way as follows. Whether the
anomalous behavior actually shows up or not also depends on the role of
repulsive core in the rotational motion of solute, which gives rise to
the collisional friction: a spherical molecule like water will be free
from the collisional friction, while a rod-like molecule such as
acetonitrile should have significant collision with solvent molecules
upon rotation. The behavior exhibited in Fig. \ref{relaxationfunction} is
consistent with the physical picture drawn above. Figs.
\ref{relaxationfunction}a) and \ref{relaxationfunction}b), which concern
water as solvent, show anomalous density dependence of the rotational
relaxation time, which disappears with increasing temperature. The
methanol-in-water system exhibits more significant anomaly compared to
the acetonitrile-in-water system. The anomalous behavior is attributed to
the strong electrostatic interaction, so called ``hydrogen bond'',
among the solvent molecules and those between solute
and solvent. The former causes the decrease in the dielectric relaxation
time with pressure, and the latter brings about the coupling between
the dielectric mode of the solvent and the rotation of the solute.
The anomaly is largely suppressed for the
acetonitrile-in-water system due to the enhanced significance of the
repulsive core in the molecule. The methanol-in-acetonitrile system does
not show any indication of the anomalous density dependence, although
there is a hydrogen bond between solute and solvent as is indicated in
Fig. \ref{gam}b). The reason is because there is strong collisional
friction on the collective reorientation of the solvent
in this case, so that the dielectric relaxation becomes slower with
pressure. The enhanced density or pressure just
amplifies the significance of the repulsive core upon rotation and
thereby of the collisional friction.

%% file: figure1.tex
\begin{figure}[!htb]
\begin{center}
\includegraphics*[bb=106 23 493 511,width=0.9\columnwidth]{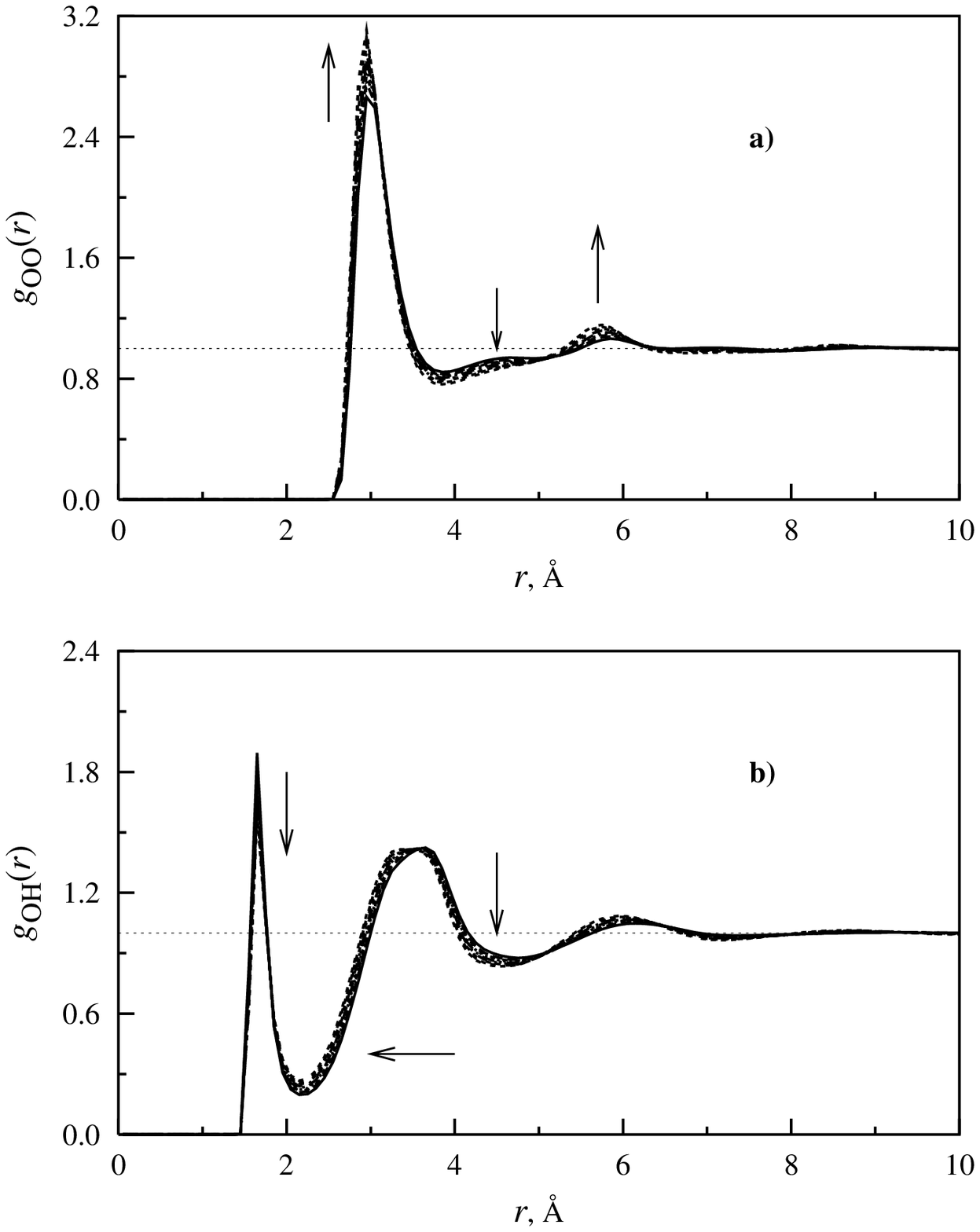}
\end{center}
\caption{Site-site radial distribution functions of neat water at
$T=273.15$ K and set of densities from 0.9 to 1.125 g/cm$^3$, obtained by
the DRISM/HNC integral equation theory. Arrows show directions of
alternations due to an increase in pressure.} \label{gww}
\end{figure}

%% file: figure2.tex
\begin{figure}[!htb]
\begin{center}
\includegraphics*[bb=105 23 493 763,width=0.9\columnwidth]{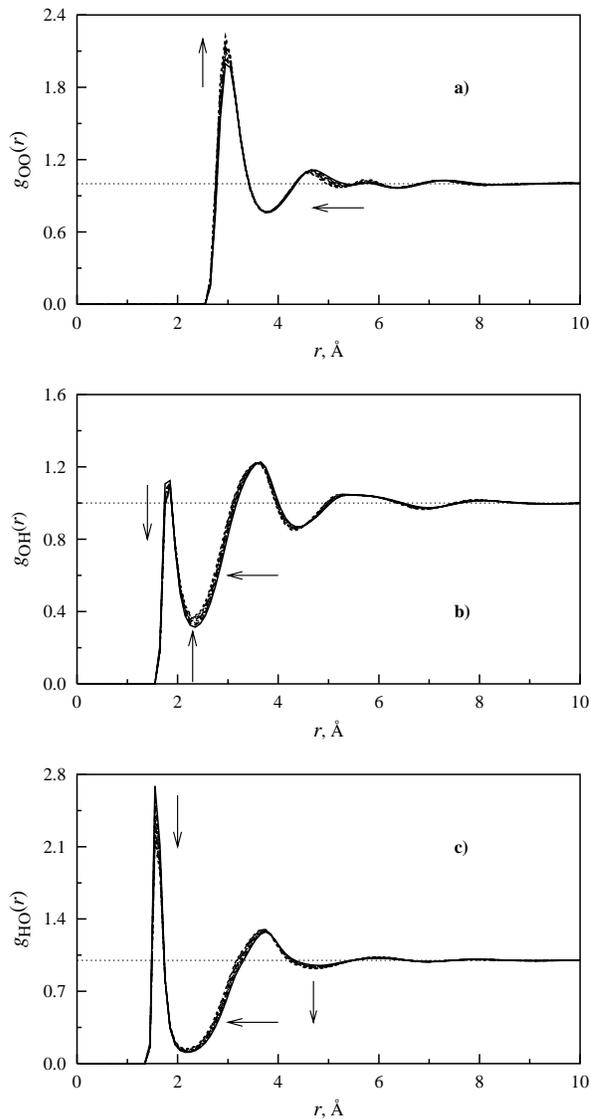}
\end{center}
\caption{Water-methanol site-site radial distribution functions for
sites, that may be related to making hydrogen bonds, at $T=273.15$ K and
set of densities from 0.9 to 1.125 g/cm$^3$ for water, obtained by the
DRISM/HNC integral equation theory. In the notations used, first site
always belongs to water, and second site always belongs to methanol.
Arrows show directions of al\-ter\-na\-ti\-ons due to an increase in
pressure.} \label{gwm}
\end{figure}

%% file: figure3.tex
\begin{figure}[!htb]
\begin{center}
\includegraphics*[bb=106 23 493 511,width=0.9\columnwidth]{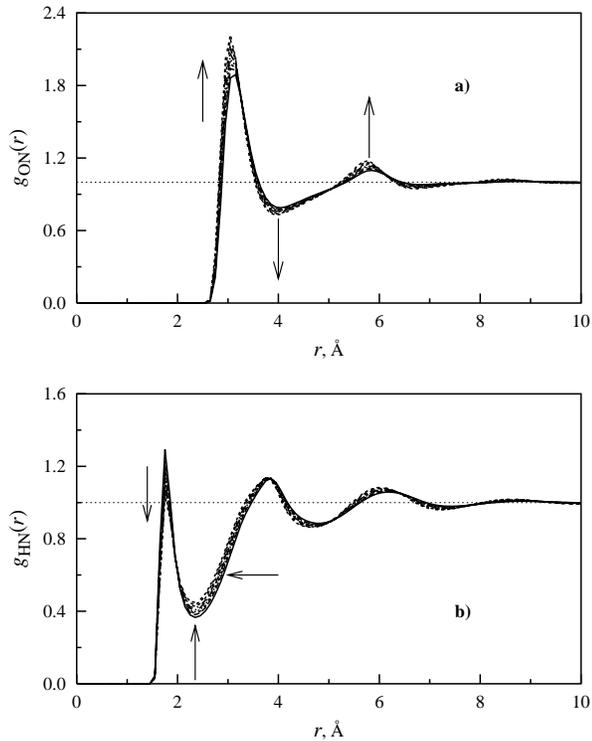}
\end{center}
\caption{Water-acetonitrile site-site radial distribution functions at
$T=273.15$ K and set of densities from 0.9 to 1.125 g/cm$^3$ for water,
obtained by the DRISM/HNC integral equation theory. In the notations
used, first site always belongs to water, and second site always belongs
to acetonitrile. Arrows show directions of al\-ter\-na\-ti\-ons due to an
increase in pressure.} \label{gwa}
\end{figure}

%% file: figure4.tex
\begin{figure}[!htb]
\begin{center}
\includegraphics*[bb=105 22 493 511,width=0.9\columnwidth]{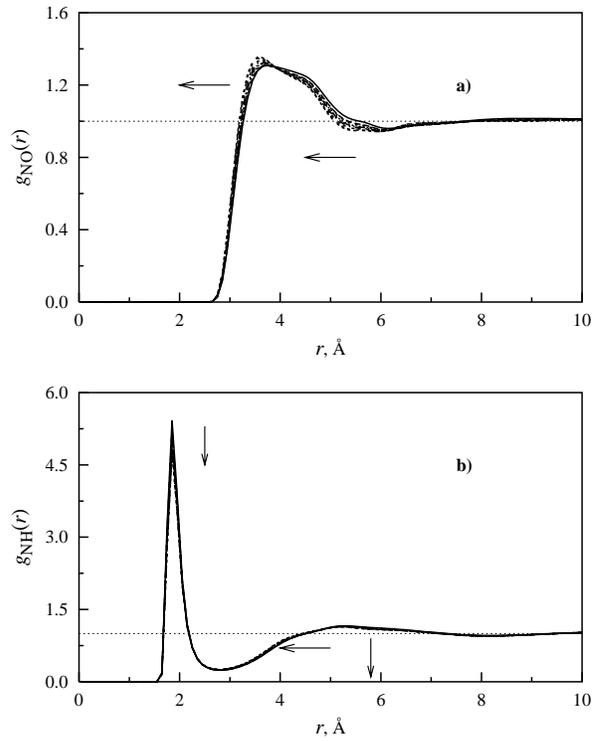}
\end{center}
\caption{Acetonitrile-methanol site-site radial distribution functions at
$T=293.15$ K and set of densities from 0.6726 to 0.815 g/cm$^3$ for
acetonitrile, obtained by the DRISM/HNC 
theory. In the notations used, first site always belongs to acetonitrile,
and second site always belongs to methanol. Arrows show directions of
alternations due to an increase in pressure.} \label{gam}
\end{figure}

%% file: figure5.tex
\begin{figure}[!htb]
\begin{center}
\includegraphics*[bb=113 21 497 763,width=0.9\columnwidth]{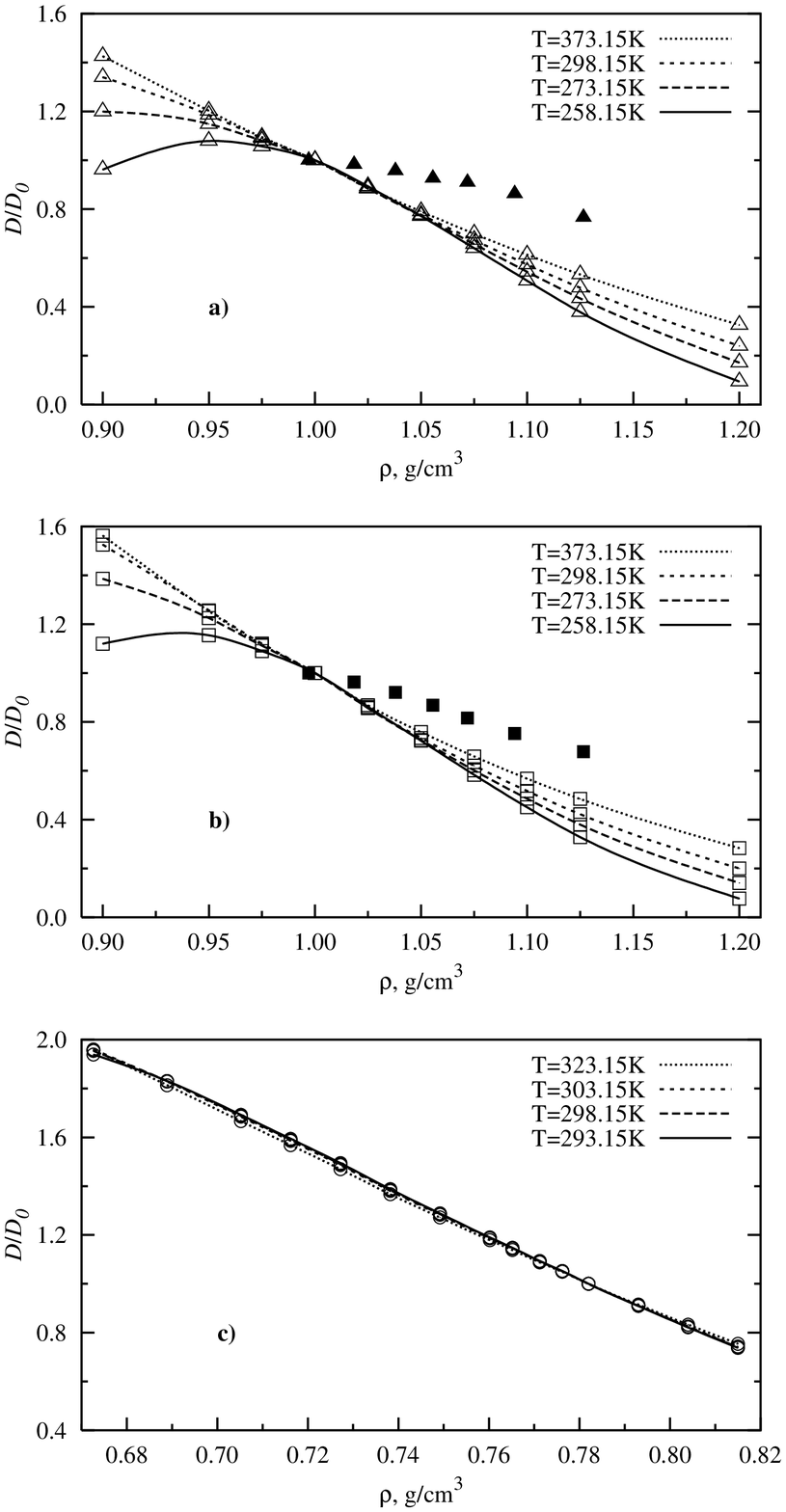}
\end{center}
\caption{Normalized translational diffusion coefficient $D/D_0$ for
methanol in water a); acetonitrile in water b); and for methaol in
acetonitrile c). {\footnotesize$\square$}, {\footnotesize$\triangle$},
{\Large$\circ$} -- theory, {\footnotesize$\blacksquare$},
$\blacktriangle$ -- results of MD simulation \protect\cite{Chowdhuri03}
for $T=298$ K, lines connecting open symbols are cubic splines for the
eye-guide.}
\label{diffusion}
\end{figure}

%% file: figure6.tex
\begin{figure}[!htb]
\begin{center}
\includegraphics*[bb=113 21 497 763,width=0.9\columnwidth]{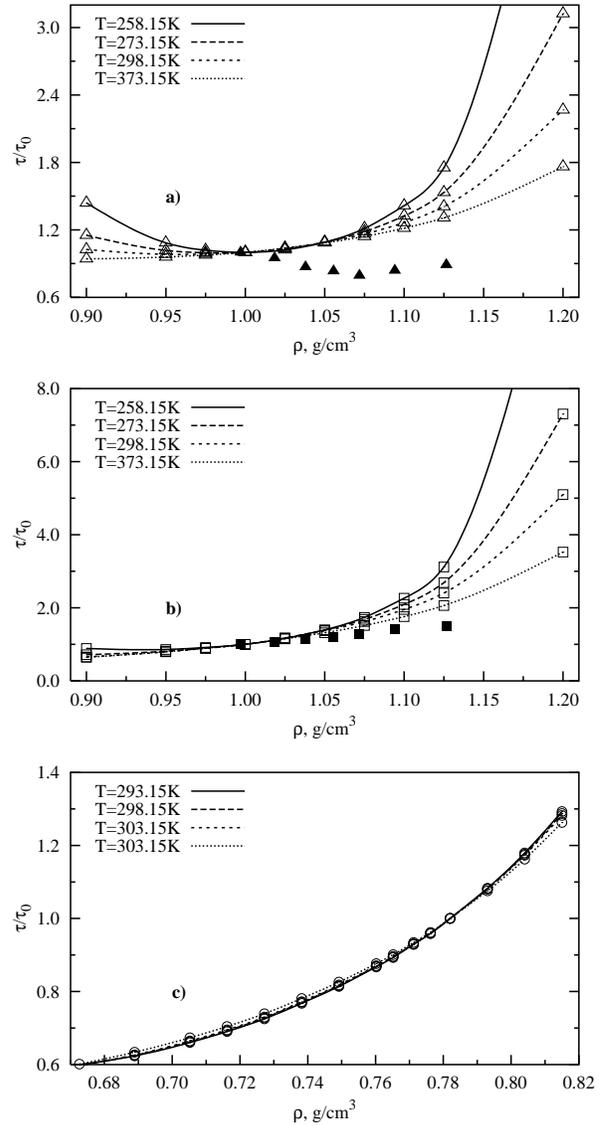}
\end{center}
\caption{Normalized reorientation relaxation time $\tau/\tau_0$ for
methanol in water a); acetonitrile in water b); and for methanol in
acetonitrile c). {\footnotesize$\square$}, {\footnotesize$\triangle$},
{\Large$\circ$} -- theory, {\footnotesize$\blacksquare$},
$\blacktriangle$ -- results of MD simulation \protect\cite{Chowdhuri03}
for $T=298$ K, lines connecting open symbols are cubic splines for the
eye-guide.}
\label{relaxationtime}
\end{figure}

%% file: figure7.tex
\begin{figure*}[!hbt]
\begin{center}
\includegraphics*[bb=18 415 576 619,width=\textwidth]{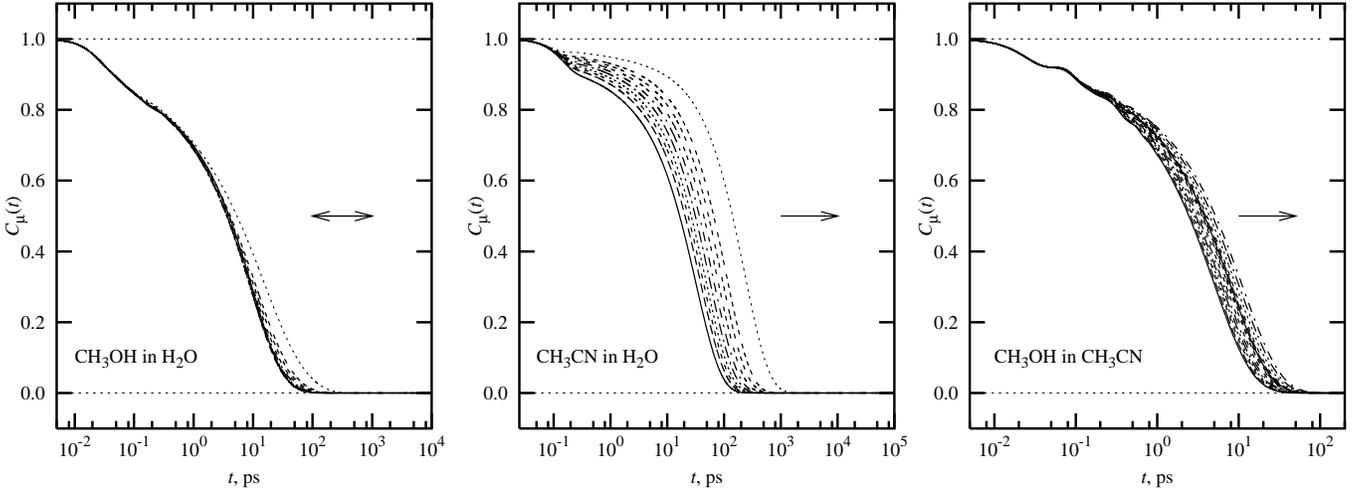}
\end{center}
\caption{Reorientation correlation functions for methanol in water (left),
acetonitrile in water (center) and methanol in acetonitrile (right) at $T=298.15$ K and
different values of density of the solvent. Arrows inside of the each plot
indicate direction(s) of the displacement for curves with the increase of density.}
\label{relaxationfunction}
\end{figure*}

%% file: figure8.tex
\begin{figure}[!htbp]
\begin{center}
\includegraphics*[bb=76 106 492 700,width=0.86\columnwidth]{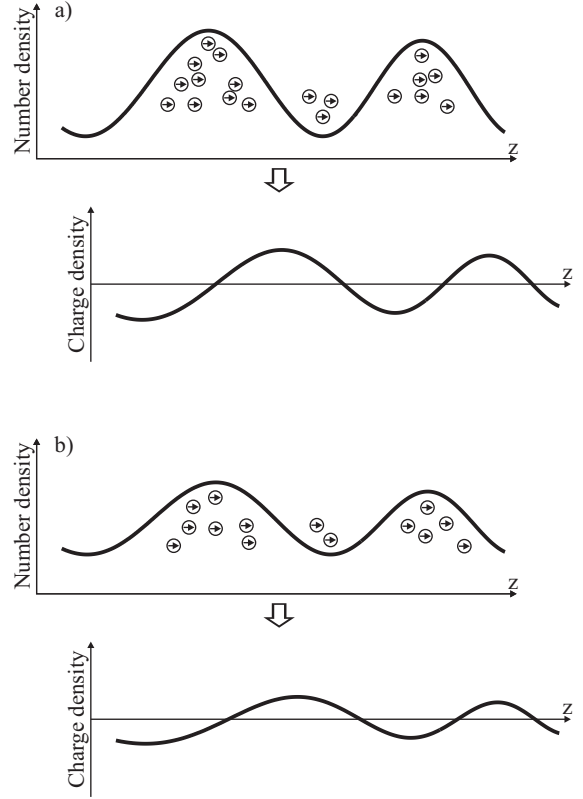}
\end{center}
\caption{Schematic view on the origin of electrostatic friction on
dielectric modes with allowance for large a) and small b) density
fluctuations.} \label{efriction}
\end{figure}

%% file: section05.tex
\section{Summary}
\label{Section05}

In present paper we have calculated the den\-si\-ty(pres\-su\-re)
dependence at various temperatures for the translational diffusion
coefficients $D$ and rank-1 reorientation relaxation times $\tau$ for
three solute-solvent systems at infinite dilution: acetonitrile and
methanol in water, and methanol in acetonitrile. Calculations have been
per\-for\-med using the site-site memory equation with the
mo\-de-coup\-ling approximation for memory kernels, and the DRISM theory
for static properties.

The theory was able to reproduce qualitatively all main features of
temperature and density dependences of $D$ and $\tau$ observed in real
and computer experiments. In particular, anomalous behavior, i.e. the
increase in mobility with density, is observed for $D$ and $\tau$ of
methanol in water at the ambient condition, while acetonitrile in water
and methanol in acetonitrile show the ordinary behavior, i.e. the
monotonous decrease in mobility with increasing density. The variety
exhibited by the different solute-solvent systems in the density
dependence of the mobility was interpreted in terms of the two
com\-pe\-ting origins of friction, which interplay with each other as
density increases:  the collisional and dielectric frictions which,
respectively, increase and decrease with increasing density.  The  cause
of the dielectric friction is the inhomogeneity in the charge density of
solvent, induced by  the number density fluctuation due to the hydrogen
bond. The suppression of the number density fluctuation and the
consequential decrease in the charge density fluctuation  are the
physical origins of the anomalous density dependence of the molecular
mobility in hydrogen-bonding systems.

%% file: section06.tex
\acknowledgments

This work is supported in part by the Grant-in-Aid for Scientific Research 
on Priority Area of ``Water and Biomolecules'' of the Japanese Ministry of 
Education, Culture, Sports, Science and Technology (MONBUKAGAKUSHO).
One of the authors (A.E.K.) expresses his gra\-ti\-tu\-de to the
\textit{Lehr\-stuhl f\"ur Ther\-mo\-dy\-na\-mik} at the
\textit{Ruhr-Uni\-ver\-si\-t\"at Bo\-hum} and personally to Prof.
Dr.-Ing. W.~Wagner for sending and letting use the software ``Basic
Package Water (IAPWS-95)'' for calculationg the thermodynamic properties
and some transport properties of H$_2$O. He also is thankful to Prof.
A.~Chandra for sending numerical data of MD simulations for acetonitrile
in water and methanol in water. All the authors thank Dr. S.-H. Chong for
his fruitful discussions.

%% file: preprint.bbl
\begin{thebibliography}{00}
\footnotesize
\enlargethispage{0.2cm}

\bibitem{Kell75}
G. S. Kell and E. Whalley,
\JCP \textbf{62}, 3496 (1975).

\bibitem{Easteal85}
A. J. Easteal and L. A. Wolf,
\JPC \textbf{89}, 1066 (1985).

\bibitem{Wakai94}
C. Wakai and M. Nakahara,
\JCP \textbf{100}, 8347 (1994).

\bibitem{Wakai97}
C. Wakai and M. Nakahara,
\JCP \textbf{106}, 7512 (1997).

\bibitem{Wakai99}
C. Wakai, N. Matubayasi, and M. Nakahara,
\JPCA \textbf{103}, 6685 (1999).

\bibitem{Harris97}
K. R. Harris and P. J. Newitt,
J. Chem. Eng. Data \textbf{42}, 346 (1997).

\bibitem{Harris98}
K. R. Harris and P. J. Newitt,
\JPCB \textbf{102}, 8874 (1998).

\bibitem{Harris99a}
K. R. Harris and P. J. Newitt,
\JPCA \textbf{103}, 6508 (1999).

\bibitem{Harris99b}
K. R. Harris and P. J. Newitt,
\JPCB \textbf{103}, 7015 (1999).

\bibitem{Biswas95}
R. Biswas, S. Roy, and B. Bagchi,
\PRL \textbf{75}, 1098 (1995).

\bibitem{Chakrabarti96}
H. Chakrabarti,
\JPCM \textbf{8}, 7019 (1996).

\bibitem{Nienhuys00}
H.-K. Nienhuys, R. A. van Santen, and H. J. Bakker,
\JCP \textbf{112}, 8487 (2000).

\bibitem{Sassi00}
P. Sassi, A. Morresi, G. Paliani, and R. S. Cataliotti,
\JPCM \textbf{12}, 3615 (2000).

\bibitem{Netz02}
P. A. Netz, F. Starr, M. C. Barbosa, and H. E. Stanley,
\JML \textbf{101}, 159 (2002).

\bibitem{Chowdhuri03}
S. Chowdhuri and A. Chandra,
\CPL \textbf{373}, 79 (2003).

\bibitem{Vargaftik96}
N. B. Vargaftik, Y. K. Vinogradov, and V. S. Yargin,
\textit{Handbook of Physical Properties of Liquids and Gases: Pure
Substances and Mixtures}, 3rd ed. (Be\-gell House, New York, 1996).

\bibitem{Eisenberg69}
D. Eisenberg and W. Kauzmann,
\textit{The Structure and Properties of Water}
(Oxford University Press, London, 1969).

\bibitem{Horne72}
\textit{Water and Aqueous Solutions}, ed. by R.A.~Hor\-ne
(Wiley, New York, 1972).

\bibitem{Franks72}
\textit{Water: A Comprehensive Treatise}, vols. 1-7.
Ed. by F. Franks (Ple\-num Press, New York, 1972-1982).

\bibitem{Born20}
M. Born,
Z. Phys. \textbf{1}, 221 (1920).

\bibitem{Debye29}
P. Debye,
\textit{Polar Molecules}
(Chemical Catalog Company, New York, 1929).

\bibitem{Onsager36}
L. Onsager,
\JACS \textbf{58}, 1486 (1936).

\bibitem{Boyd61}
R. H. Boyd,
\JCP \textbf{35}, 1281 (1961).

\bibitem{Zwanzig63a}
R. Zwanzig,
\JCP \textbf{38}, 1603 (1963).

\bibitem{Zwanzig63b}
R. Zwanzig,
\JCP \textbf{38}, 1605 (1963).

\bibitem{Zwanzig70a}
R. Zwanzig,
\JCP \textbf{52}, 3625 (1970).

\bibitem{Zwanzig70b}
T.-W. Nee and R. Zwanzig,
\JCP \textbf{52}, 6535 (1970).

\bibitem{Zwanzig74}
C.-M. Hu and R. Zwanzig,
\JCP \textbf{60}, 4354 (1974).

\bibitem{Hubbard77}
J. B. Hubbard and L. Onsager,
\JCP \textbf{67}, 4850 (1977).

\bibitem{Hubbard78a}
J. B. Hubbard,
\JCP \textbf{68}, 1649 (1978).

\bibitem{Hubbard78b}
J. B. Hubbard and P. G. Wolynes,
\JCP \textbf{69}, 998 (1978).

\bibitem{Zwanzig85}
R. Zwanzig and A. K. Harrison,
\JCP \textbf{83}, 5861 (1985).

\bibitem{Hirata03}
\textit{Molecular Theory of Solvation}, ed. F. Hirata.
In series: Understanding Che\-mi\-cal Reactivity, vol. 24,
series editor P. G. Mezey (Kluwer, Dordrecht, 2003).

\bibitem{Hirata81}
F. Hirata and P. J. Rossky,
\CPL \textbf{83}, 329 (1981).

\bibitem{Hirata82}
F. Hirata, B. M. Pettitt and P. J. Rossky,
\JCP \textbf{77}, 509 (1982).

\bibitem{Hirata83}
F. Hirata, P. J. Rossky and B. M. Pettitt,
\JCP \textbf{78}, 4133 (1983).

\bibitem{Mori65a}
H. Mori,
\PTP \textbf{33}, 423 (1965).

\bibitem{Mori65b}
H. Mori,
\PTP \textbf{34}, 399 (1965).

\bibitem{Zubarev9697}
D. N. Zubarev, V. G. Morozov, G. R\"opke,
\textit{Statistical Mechanics of Non-Equilibrium Processes}.
Vol. 1: \textit{Basic Concepts, Kinetic Theory}
(Springer, Berlin, 1996), and
Vol. 2: \textit{Relaxation and Hydrodynamical Processes}
(Springer, Berlin, 1997).

\bibitem{Ferziger72}
J. H. Fer\-zi\-ger and H. G. Ka\-per,
\textit{Ma\-the\-ma\-ti\-cal The\-o\-ry of Trans\-port Pro\-ces\-ses in Ga\-ses}
(North-Hol\-land, Ams\-ter\-dam, 1972).

\bibitem{Hirata92}
F. Hirata,
\JCP \textbf{96}, 4619 (1992).

\bibitem{Madden82}
P. Madden and D. Kivelson,
\JPC \textbf{86}, 4244 (1982).

\bibitem{Calef89}
D. F. Calef and P. G. Wolynes,
\JCP \textbf{78}, 4145 (1983).

\bibitem{Chandra89}
A. Chandra and B. Bagchi,
\JCP \textbf{91}, 1829 (1989).

\bibitem{Wei89}
D. Wei and G. N. Patey,
\JCP \textbf{91}, 7113 (1989).

\bibitem{Wei90}
D. Wei and G. N. Patey,
\JCP \textbf{93}, 1399 (1990).

\bibitem{Hirata88}
F. Hirata, P. Redfern, and R. M. Levy,
Int. J. Quant. Chem. \textbf{15}, 179 (1988).

\bibitem{Chong99}
S.-H. Chong and F. Hirata,
\JCP \textbf{111}, 3654 (1999).

\bibitem{Hirata95}
F. Hirata, T. Munakata, F. Raineri, and H. L. Friedman,
\JML \textbf{65}-\textbf{66}, 15 (1995).

\bibitem{Chong03}
In Ref. \onlinecite{Hirata03}, Chapter 5:
\textit{Dynamical Processes in Solution}, by S.-H.~Chong, p. 277-349.

\bibitem{Yamaguchi01a}
T. Yamaguchi, N. Matubayasi, and M. Nakahara,
\JCP \textbf{115}, 422 (2001).

\bibitem{Yamaguchi01b}
T. Yamaguchi and F. Hirata,
\JCP \textbf{115}, 9340 (2001).

\bibitem{Yamaguchi02a}
T. Yamaguchi, S.-H. Chong, and F. Hirata,
\JCP \textbf{116}, 2502 (2002).

\bibitem{Yamaguchi02b}
T. Yamaguchi and F. Hirata,
\JCP \textbf{117}, 2216 (2002).

\bibitem{Yamaguchi03a}
T. Yamaguchi, S.-H. Chong, and F. Hirata,
\JCP \textbf{119}, 1021 (2003).

\bibitem{Yamaguchi03b}
T. Yamaguchi, A. Nagao, T. Matsuoka, and S. Koda,
\JCP \textbf{119}, 11305 (2003).

\bibitem{Yamaguchi03c}
T. Yamaguchi, S.-H. Chong, and F. Hirata,
\MP \textbf{101}, 1211 (2003).

\bibitem{Yamaguchi04a}
T.~Yamaguchi, T.~Matsuoka, and S.~Koda,
\JCP \textbf{120}, 7590 (2004).

\bibitem{Yamaguchi04b}
T. Yamaguchi, S.-H. Chong, and F. Hirata,
\JML \textbf{112}, 117 (2004).

\bibitem{Nishiyama03}
K. Nishiyama, F. Hirata, and T. Okada,
\JCP \textbf{118}, 2279 (2003).

\bibitem{Chong98c}
S.-H. Chong and F. Hirata,
\PRE \textbf{58}, 7296 (1998).

\bibitem{Berne70}
B. J. Berne and G. D. Harp,
in: Advances in Chemical Physics, vol. XVII, p. 63-227.
Ed. by I. Prigogine and S. Rice
(Intercsience Publishers, New York, 1970).

\bibitem{Hansen86}
J. P. Hansen and I. R. McDonald,
\textit{Theory of Simple Liquids}, 2nd ed.
(Academic Press, London, 1986).

\bibitem{Perkyns92a}
J. Perkyns and M. B. Pettitt,
\CPL \textbf{190}, 626 (1992) 626.

\bibitem{Perkyns92b}
J. Perkyns and M. B. Pettitt,
\JCP \textbf{97}, 7656 (1992).

\bibitem{Chong98b}
S.-H. Chong and F. Hirata,
\PRE \textbf{58}, 6188 (1998).

\bibitem{Chong02}
S.-H. Chong and W. G\"otze,
\PRE \textbf{65}, 41503 (2002).

\bibitem{Berendsen87}
H. J. J. C. Berendsen, J. R. Grigera, and T. P. Straatsma,
\JPC \textbf{91}, 6269 (1987).

\bibitem{Edwards84}
D. M. Edwards, P. A. Madden, and I. R. McDonald,
\MP \textbf{51}, 1141 (1984).

\bibitem{Jorgensen86}
W. L. Jorgensen,
\JPC \textbf{90}, 1276 (1986).

\bibitem{Wagner02}
W. Wagner and A. Pru\ss,
J. Chem. Phys. Ref. Data \textbf{31}, 387 (2002).

\bibitem{LandoltBornstein80}
\textit{Landolt-B\"ornstein Numerical Data and Functional Relationships
in Science and Technology}, New Series, ed. in chief K.-H. Hellwege.
Group IV: Macroscopic and Technical Properties of Matter. Vol.~4:
High-Pressure Properties of Matter. G. Beggerow. Ed. Kl. Sch\"afer
(Springer-Verlag, Berlin, 1980).

\bibitem{Ku98}
H.-C. Ku and C.-H. Tu,
J. Chem. Eng. Data \textbf{43}, 465 (1998).

\bibitem{Nikam98}
P. S. Nikam, L. N. Shirsat and M. Hasan,
J. Chem. Eng. Data \textbf{43}, 732 (1998).

\bibitem{Delta}
Dielectric constants data from ``Delta Construction Corporation'',
www.deltacnt.com.

\bibitem{ASI}
Dielectric constants data from ``ASI Instruments Inc.'',\newline
www.asiinstr.com.

\bibitem{Cavell65}
E. A. S. Cavell, H. G. Jerrard, B. A. W. Sim\-monds, and J. A. Spe\-ed,
\JPC \textbf{69}, 3657 (1965).

\bibitem{Kovalenko99}
A. Kovalenko, S. Ten-no, and F. Hirata,
J. Comp. Chem. \textbf{20}, 928 (1999).

\bibitem{Luzar00}
A. Luzar,
\JCP \textbf{113}, 10663 (2000).

\bibitem{Mezei81}
M. Mezei, D. L. Beveridge,
\JCP \textbf{74}, 622 (1981).

\bibitem{Rahman71}
A. Rahman and F. H. Stillinger,
\JCP \textbf{55}, 3336 (1971).

\bibitem{Yoshida02}
K. Yoshida, T. Yamaguchi, A. Kovalenko, and F. Hirata,
\JPCB \textbf{106}, 5042 (2002).

\end{thebibliography}
